\documentclass{article}

\usepackage{PRIMEarxiv}

\usepackage[utf8]{inputenc} % allow utf-8 input
\usepackage[T1]{fontenc}    % use 8-bit T1 fonts
\usepackage{hyperref}       % hyperlinks
\usepackage{url}            % simple URL typesetting
\usepackage{booktabs}       % professional-quality tables
\usepackage{amsfonts}       % blackboard math symbols
\usepackage{nicefrac}       % compact symbols for 1/2, etc.
\usepackage{microtype}      % microtypography
\usepackage{lipsum}
\usepackage{fancyhdr}       % header
\usepackage{graphicx}       % graphics
\graphicspath{{media/}}     % organize your images and other figures under media/ folder

\title{Social Media COVID-19 Contact Tracing
Using Mobile Social Payments and
Facebook Data}
\author{
  Shrivu Shankar \\
  Computational Media Lab \\
  Department of Computer Science \\
  The University of Texas at Austin \\
  \texttt{shrivu.shankar@utexas.edu} \\
  \And
  Dhiraj Murthy \\
  Computational Media Lab \\
  School of Journalism and Media \\
  Moody College of Communication \\
  The University of Texas at Austin \\
  \texttt{Dhiraj.Murthy@austin.utexas.edu} \\
  \And
  Hassan Dashtian \\
  Computational Media Lab \\
  The University of Texas at Austin \\
  \texttt{dashtian@utexas.edu} \\
}

\begin{document}
\maketitle

\begin{abstract}
Many in the US were reluctant to report their COVID-19 cases at the height of the pandemic (e.g., for fear of missing work or other obligations due to quarantine mandates). Other methods such as using public social media data can therefore help augment current approaches to surveilling pandemics. This study evaluated the effectiveness of using social media data as a data source for tracking public health pandemics. There have been several attempts at using social media data from platforms like Twitter for analyzing the COVID-19 pandemic. While these provide a multitude of useful insights, new platforms like Venmo, a popular U.S. mobile social payment app often used during in-person activities, remain understudied. We developed unique computational methods (combining Venmo- and Facebook- derived data) to classify post content, including the location where the content was likely posted. This approach enabled geotemporal COVID-19-related infoveillance. By examining 135M publicly available Venmo transactions from 22.1M unique users, we found significant spikes in the use of COVID-19 related keywords in March 2020. Using Facebook-based geotags for 9K users along with transaction geo-parsing (i.e., parsing text to detect place names), we identified 38K location-based clusters. Within these groups, we found a strong correlation (0.81) between the use of COVID-19 keywords in a region and the number of reported COVID-19 cases as well as an aggregate decrease in transactions during lockdowns and an increase when lockdowns are lifted. Surprisingly, we saw a weak negative correlation between the number of transactions and reported cases over time (-0.49). Our results indicate that using non-Twitter social media trace data can aid pandemic- and other health-related infoveillance.
\end{abstract}

% keywords can be removed

\section{Introduction}
As of January 2023, there have been nearly 673M cases and 6.9M dead worldwide due to COVID-19 \cite{Dong2020-re}. Pursuing novel methods of data collection is important for both the COVID-19 pandemic as well as future ones. As 85\% of U.S. adults own a smartphone \cite{Sun2023-ct}, mobile trace data can be used  to improve the scale and accuracy of existing infoveillance methods for spread analysis and contact tracing. However, installing privacy-considerate contact tracing software on these phones is non-trivial and poses extremely difficult challenges in countries where there is a high level of distrust of government or regulations that already oppose these types of data collection. Less than half of surveyed U.S. adults endorsed contact-tracing apps \cite{Zhang2020-mm} yet more than 72\% report using a social media platform \cite{auxier2021social}. This begs the question: Can social media activity provide an effective alternative form of infoveillance?

While current work in social media-based infoveillance is  generally accomplished using data from Twitter, we performed a similar analysis on social payment transactions, which are known to include shorter text but more likely to be location-based \cite{acker2020venmo}. In this study, we evaluated whether Venmo transactions, which include mandatory, short ‘memo’ messages between users sending or requesting money, a passive data source, are useful for tracking a pandemic. Using memos rather than tweets introduced an added challenge of deriving data from shorter text which is mediated by cross-referencing data with accounts on Facebook. Combining Venmo and Facebook data helps to acquire more information about the transactions. During the height of the pandemic the platform had 40M users \cite{acker2020venmo}. By 2023, Venmo had over 70 million users \cite{curry2021venmo}. Venmo allows users to send or request money with short personalized messages attached which are by default publicly available\footnote{\href{https://help.venmo.com/hc/en-us/articles/210413717-Payment-Activity-Privacy}{https://help.venmo.com/hc/en-us/articles/210413717-Payment-Activity-Privacy}}. These transactions often occur after or during physical interaction between users \cite{drenten2022digital}, making it a potentially more robust platform for estimating an interaction model than other social media such as Twitter.  Importantly, Venmo transactions also tend to occur among groups of friends who interact with each other (even if the transaction itself is done remotely) \cite{drenten2022digital} which is not the case on Twitter and other social media sites.  

Using Venmo’s public API, we collected 135M transactions for 22.1M users. Some users mark their transactions as private and they are only visible between the users involved in the transaction; as these data are not public, they were not included in our data set. We found significant spikes in the use COVID-19 related keywords such as “quarantine”, “covid”, “social distancing”, and “cough” in March 2020. By combining our data with Facebook-based geotags for 9K users along with transaction geo-parsing, we estimate and identify 38K location-based clusters. With these groups, we found a strong correlation between the use of COVID-19 keywords in a region and the number of reported COVID-19 cases (0.81). Moreover, we found that there was an aggregate decrease in transaction frequency during lockdowns as well as an increase when lockdowns are lifted. We unexpectedly observed a weak negative correlation between the number of transactions and reported cases over time (-0.49). Our results indicate that alternative (i.e., non-Twitter) social media-based data is promising for providing pandemic-related insights, especially among populations with low access to traditional forms of health information, but high social media usage. Social media-based location clusters derived from less studied platforms can be used in conjunction with other data sources to help curtail both the current COVID-19 pandemic as well as future pandemics.

\section{COVID-19 Location-Based Clustering}
Several studies have demonstrated the effectiveness of location-based clustering as a method of curbing the spread of a viral outbreak (McCloskey et al 2017), including research during the COVID-19 pandemic (Firth et al 2020). Previous work has used survey data and simulations to show the utility of location-based clustering in curtailing the spread of COVID-19 or misinformation related to COVID-19 \cite{chang2021mobility}. This work found that while contact tracing can help thwart an epidemic, logistical data collection challenges prevent it from being fully effective. In addition, the effectiveness of contact tracing may change as a result of intrinsic epidemiological properties and curtailment strategies. Moreover, there are no globally standardized approaches for contact tracing. Instead, several governments and organizations have developed applications aimed at collecting contact data and actioning it effectively for their specific constituents\footnote{\href{https://en.wikipedia.org/wiki/COVID-19_apps}{https://en.wikipedia.org/wiki/COVID-19\_apps}}. Among these are Singapore’s Trace Together app, the UK’s NHS COVID-19 contact tracing app, France’s StopCOVID app, South Korea’s public COVID database, and China’s contact tracing integration with WeChat and Alipay \cite{alanzi2021review}.

While these apps have likely helped slow the spread of the virus, privacy and ethics questions around their use have been the subject of government and various public debates \cite{lucivero2022normative}. Several studies uncover how this is a complex privacy issue that flares around questions regarding what data should be collected, who should collect it, and how it should be used. Despite these challenges, several proof-of-concept COVID-19 app-augmented contact tracing studies have been published. One voluntary program within WeChat (a primarily Chinese-used app with over one billion active users) allows users to submit their GPS-based location, time, and diagnosis \cite{wang2020new}. This data is then used to compute a “risk index” that can be used to notify individuals who should quarantine. The PrivateKit app aims to address concerns around security, privacy, and abuse through fully consensual tracking \cite{yasaka2020peer}. While taking significant privacy precautions, PrivateKit is still limited by its need to store location data and government reliance.

Although there are risks in using social media data for disease control, public backlash against contact tracing apps means that in countries like the U.S., social media-based infoveillance and location-based clustering methods could be an incredibly useful (and publicly acceptable) alternative. Furthermore, in lieu of collecting true physical contact data, using other features such as communication traces can be invaluable inputs to infoveillance efforts. Through numerical simulations, Farrahi et al. show that while there was not a significant overlap between communication traces and contact traces, the data known could still be used to significantly reduce the total number of infected cases \cite{farrahi2014epidemic}. Moreover, machine learning models have been shown to successfully discover discussion of COVID related symptoms among Twitter users based on the content of their tweets. However, the question remains of whether social media posts by users (as communication traces were used in Haggag et al 2021 \cite{Haggag2021-iz}) could be used as a data source for contact tracing on more niche platforms such as Venmo.

\section{Social Media Geolocation}
Social media geolocation involves discovering geographical locations and can often be used with statistical methods to make correlations with government data. Social media location prediction research is often done on Twitter due to its popularity and dataset availability. Zheng et al. \cite{zheng2018survey} provides an overview of the location prediction methods that have been employed on Twitter so far. They separate the problem into home, tweet, and mentioned location prediction as well as provide an overview of methods used including geolocation based on friendships, social-closeness, time posting, and/or keywords \cite{zheng2018survey}. Ground truth data for geolocation studies often comes from geotagged tweets and GEOTEXT (and/or derived) datasets. Of the methods analyzed, social-closeness (location inference based on the contact between users) and friendship-based (location inference based on digital friendships) methods are the most robust for contract tracing as features can be derived from transactions between users. After testing both “end-to-end locative expression recognisers” and “geospatial named entity recognizers” on a variety of corpuses Liu et al. (2014) find StanfordNER\footnote{\href{https://nlp.stanford.edu/ner/}{https://nlp.stanford.edu/ner/}} to be the best performing on a tweet-based corpus \cite{liu2014automatic}.

\section{Previous Research on Venmo}
Since its inception in 2009 and its creation of an API for accessing the transaction history of its users, Venmo has been the topic of several research studies, which seek to examine the behavior and characteristics of its users. For example, previous work has used the Venmo API to create a social graph of several million users. Specifically, Zhang et al. \cite{zhang2017cold} and Unger et al. \cite {unger2020examining} were able to identify unique transaction patterns of business and users, dense transaction-based communities, and the existence of “niche groups” where a community pertains to only a specific type of transaction. They also discovered undocumented characteristics of the Venmo API that make it easier to index and that 3.1\% of transactions exhibited recurring periodicity. Moreover, Unger et al. replicate and confirm many of the observations still held years later, although they also found an increased number of small communities as well as an increase in single-transaction users \cite{unger2020examining}. A study using interviews with Venmo users found that they increasingly see Venmo as a social platform and, generally, those interviewed did not lessen their Venmo use based on the public sharing of their transactions. Others have conceptualized Venmo as a social platform and studied how users are sharing and interacting via their transactions, particularly through emoji trends, temporal patterns, and keyword usage \cite{murthy2020understanding}.

By combining Venmo transaction data with data from the Facebook API, Squire \cite{squire2019understanding} analyzes the social and organizational structure of the “Proud Boys” grey network, which “conduct[s] a mixture of legal and illegal activities” and has an organizational structure that may be only partially known. After indexing confirmed Proud Boy accounts, they were able to determine members and geographically tie organization “Elders” (i.e., leaders) and members to cities as well as analyze the relationships between each role. This and other work has found that the latent attributes in Venmo transaction notes make for a rich data source \cite{stobaugh2023predicting}. With two million users and ground truth seed locations found by cross-referencing identities with geotagged Facebook profiles, Yao \cite{yao2018beware} achieves 90\% top-5 accuracy in predicting where a Venmo user is located. 

\section{Methods}
\subsection{Collecting Venmo Data}
To collect our dataset of Venmo transactions, we use a cluster of web scrapers to query Venmo’s publicly available transactions-by-users API. Previous work found that Venmo IDs are sequential, which allows us to query the API by querying every ID between one and the total number of users\cite{zhang2017cold}. We consolidate and deduplicate by inserting discovered users and transactions into a SQL database. By strategically sampling the API, we found that the last and latest user ID, at the time of writing, is 95,251,463 (implying there are 95M total Venmo users; though some of these IDs likely represent inactive or deleted accounts). As rate limiting is done via Venmo’s content delivery network (CDN), Amazon CloudFront\footnote{\href{https://aws.amazon.com/cloudfront/}{https://aws.amazon.com/cloudfront/}}, and that request rate limits are enforced on a per-IP basis, we horizontally scaled our data collection jobs on our campus supercomputing cyberinfrastructure by sharding the range of user ids across up to 20 machines (total CPU hours = 120 hours) and employing the use of HTTP proxy servers to increase our range of unique IP addresses. During our initial collection period, we were able to scrape around 9M transactions per IP per day.

\subsection{Inferring User Locations}
We use Yao’s \cite{yao2018beware} method of establishing ground truth user locations based on matching Facebook accounts to Venmo using profile picture similarity\footnote{We contacted Dr. Yao, but did not receive their data nor confirmation on how they extracted data from Facebook; we therefore extrapolated methods from what was described in their published work.}. We continuously sample users with non-default profile pictures and use their first and last name to search for Facebook profiles on Bing and directly on Facebook. For every resulting profile, we compute the image similarity of the profile’s picture with the picture of the user on Venmo (included in the Venmo API response). Profile pictures with high image similarity are considered to be the same user. We found the profile picture URLs returned by the Venmo API occasionally include the suffix “\&facebook=true”. While undocumented, this likely means the picture was uploaded through an integration with Facebook and we filter by the existence of this when sampling users. Once Venmo users are paired directly with a Facebook profile, we use the web browser automation framework Selenium\footnote{\href{https://www.selenium.dev/}{https://www.selenium.dev/}} to visit those public profiles and extract geotags in text, such as“Lives in ...” or “From ...”.

\begin{figure}[hbt!]
  \centering
  \includegraphics[width=0.55\columnwidth]{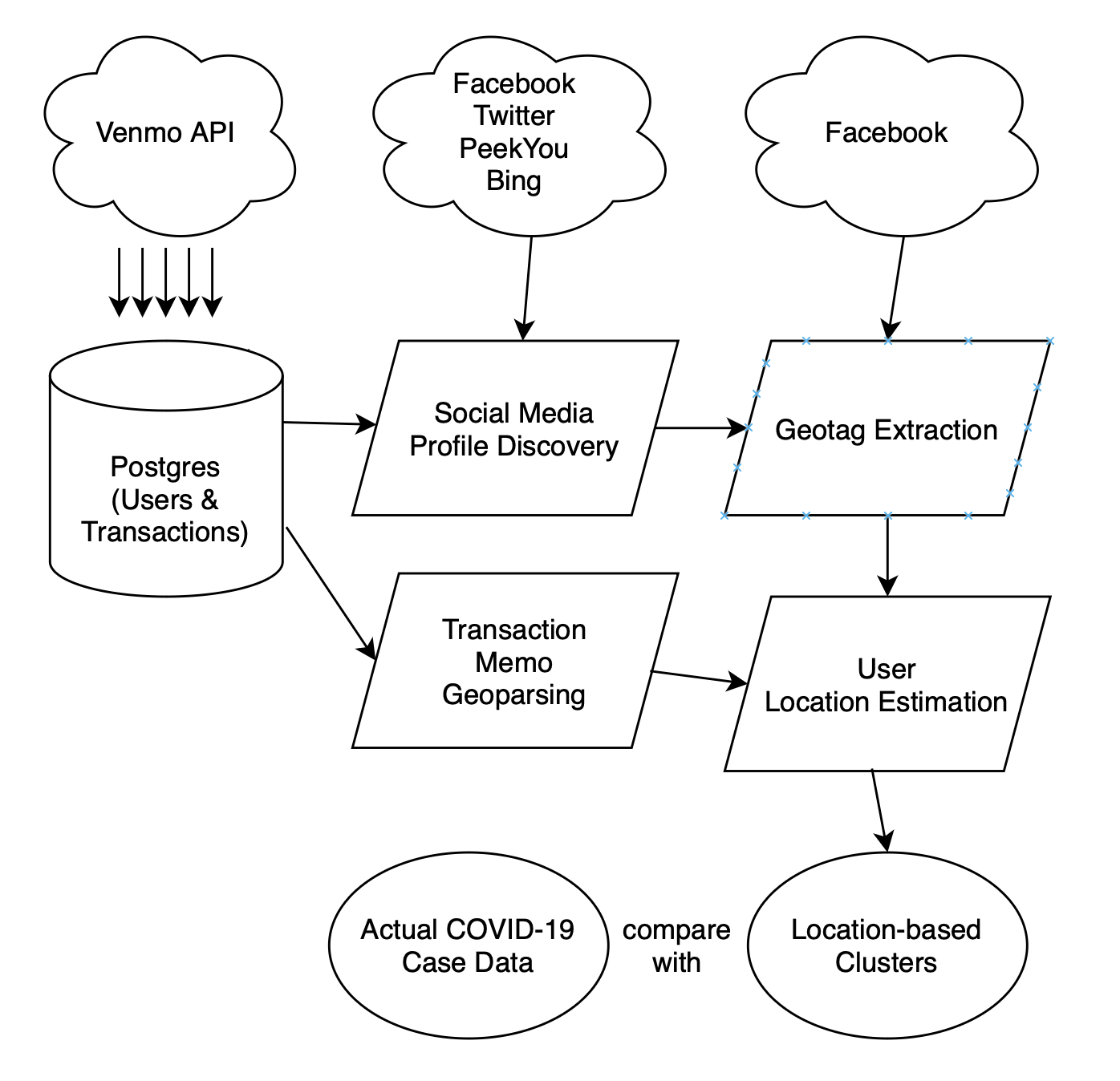}
  \caption{An overview of our data processing methods.}
  \label{flowchart}
\end{figure}

In our pilot testing, we found that Facebook’s rate limiting created a significant bottleneck, with only tens of profiles successfully being queried and matched before requests were rate limited. Since the queries require authentication, scaling across several machines was not an option. Another method we evaluated was using the raw social media profiles found from Bing results (profiles in which we have not confirmed that the profile pictures match; only their names do) for Venmo users who have unique names in the dataset as this would not be as constrained by rate limiting and does not require authentication. Such an approach, if successful, would allow a larger portion of our dataset to be geolocated. However, we found this method would frequently yield several matches for each user from distinct locations with no additional information to disambiguate. To resolve this, we used PeekYou\footnote{\href{https://www.peekyou.com/}{https://www.peekyou.com/}}, a 3rd party social media search engine, to provide both a way of enumerating public accounts under a user’s name, but also their profile pictures for comparison with profile pictures used in Venmo. We found that PeekYou has no significant rate limiting restraints and can be used in parallel. Once PeekYou Facebook profile matches were found, we used a Selenium-based Facebook scraper for the final step of discovering geotags on matched profiles. Our data methods are illustrated in Figure \ref{flowchart}. Our data collection and inferential methods were submitted and approved by our University’s Institutional Review Board (IRB).

\subsection{Discerning Location-Based Clusters}
To create an interaction graph, we make every user a vertex and connect users with an edge if they have had at least four transactions between them (from March to October 2020). Using connected components, we segmented the graph into a set of clusters, who we estimate could have had physical contact and could have thus transmitted the virus between each other if infected. Since the number of known user locations is limited, we use the geoparsing library Mordecai on transaction messages between users to estimate their geolocation. To disambiguate groups that include several geographically related words, we pick the most frequent U.S. location mentioned as this tended to match with the results of geotagged users. For each of these groups, we then use those geoparsed keywords and known user locations from Facebook to find and estimate a geographical position. These geolocated groups are then compared with the known COVID-19 statistics in those locations. We also examine the use of COVID-19 related phrases (e.g., “self-isolating”) in these groups.

\section{Results}
\subsection{Data Collection}
Through the Venmo API, we were able to obtain the data of 22.1M users and 136M transactions. Of these users, 2.4M (11\%) had an identifiable “facebook=true” tag on their profile, signifying that the account was integrated with Facebook. Focusing on these users, we were able to collect Bing profile search results for all 2.4M users, Facebook profile search results for 23K users, and PeekYou profile search results for 393K users. While both the Facebook and PeekYou results contained profile pictures that can be used to match Facebook and Venmo profiles, our ability to collect more data from PeekYou led us to primarily use that for matching. Of our PeekYou query results, we were able to find a high image similarity with at least one returned profile to a user’s Venmo profile for 19K users. Of these users, we were able to directly obtain from Facebook geotags for 7K users (0.3\% of Venmo users with Facebook integration). In theory, nearly all 19K users could have been geolocated if Facebook did not employ rate limiting measures during data collection. We also suspect that many of the Facebook profile pictures returned by PeekYou were outdated, which would have led to a higher false-negative rate when matching profiles than would be found if the profile pictures were collected directly from Facebook.

\subsection{Inferring User Locations}
\begin{figure}[hbt!]
  \centering
  \includegraphics[width=0.75\columnwidth]{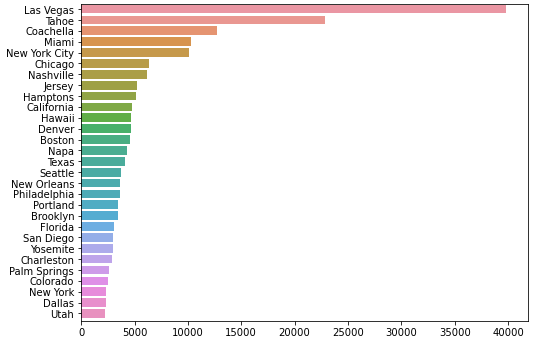}
  \caption{The top location keywords determined by the Python geoparsing package Mordecai in Venmo transaction messages (ordered by frequency). Non-U.S.locations are removed.}
  \label{mordecai}
\end{figure}

We first look at the results of geoparsing transaction messages as this could be used to geo-locate a larger portion of the dataset than can be done with Facebook profile geotags. We found an average of 1 geographically related token per 40 transactions. As indicated by Figure \ref{mordecai}, these tokens seem to be used most often for transactions that take place while the user is traveling. Rather than discovering where people live, we often identified where people are vacationing. For example, the top 3 tokens we identified (Las Vegas, Tahoe, and Coachella) are primarily U.S. tourist destinations. This conclusion is compounded by the results of Table \ref{geographical_tokens}, which provides a random sample of how these top 3 tokens are used. The names of these locations with “hotel”, “car + gas”, and “snacks”, indicate the transactions were done as a part of the user traveling with other users. We therefore assume that mentioning one’s home location in these phrases is unlikely. There were also several references to locations outside of the U.S., often coming from users using the name of a foreign country or city as an adjective such as “China town” although this was not common among references to U.S. locations. It is also noted that the frequencies found in Figure \ref{mordecai} seem to follow Zipf’s Law \cite{reed2001pareto} (i.e., the occurrence of a word is inversely proportional to its rank position in a frequency table) which could be an illustration of “preferential attachment”, ‘in which a new node has a higher probability to attach to one of the most connected ones’ among tourist destination \cite{baggio2008network} such that the most mentioned places on Venmo inspire other users to visit and Venmo while there in a self-reinforcing cycle. While the results of Mordecai are accurate in terms of identifying geographically related tokens, the use of these tokens makes it difficult to rely on this as a true indicator of where users physically interact longer-term. This furthers the need for better ground truth geodata such as the Facebook geotag scraping done by Yao \cite{yao2018beware}. This difficulty of estimating home locations via geoparsing has also been found in Twitter-related research \cite{zheng2018survey}.

\begin{figure}[hbt!]
  \centering
  \includegraphics[width=0.65\columnwidth]{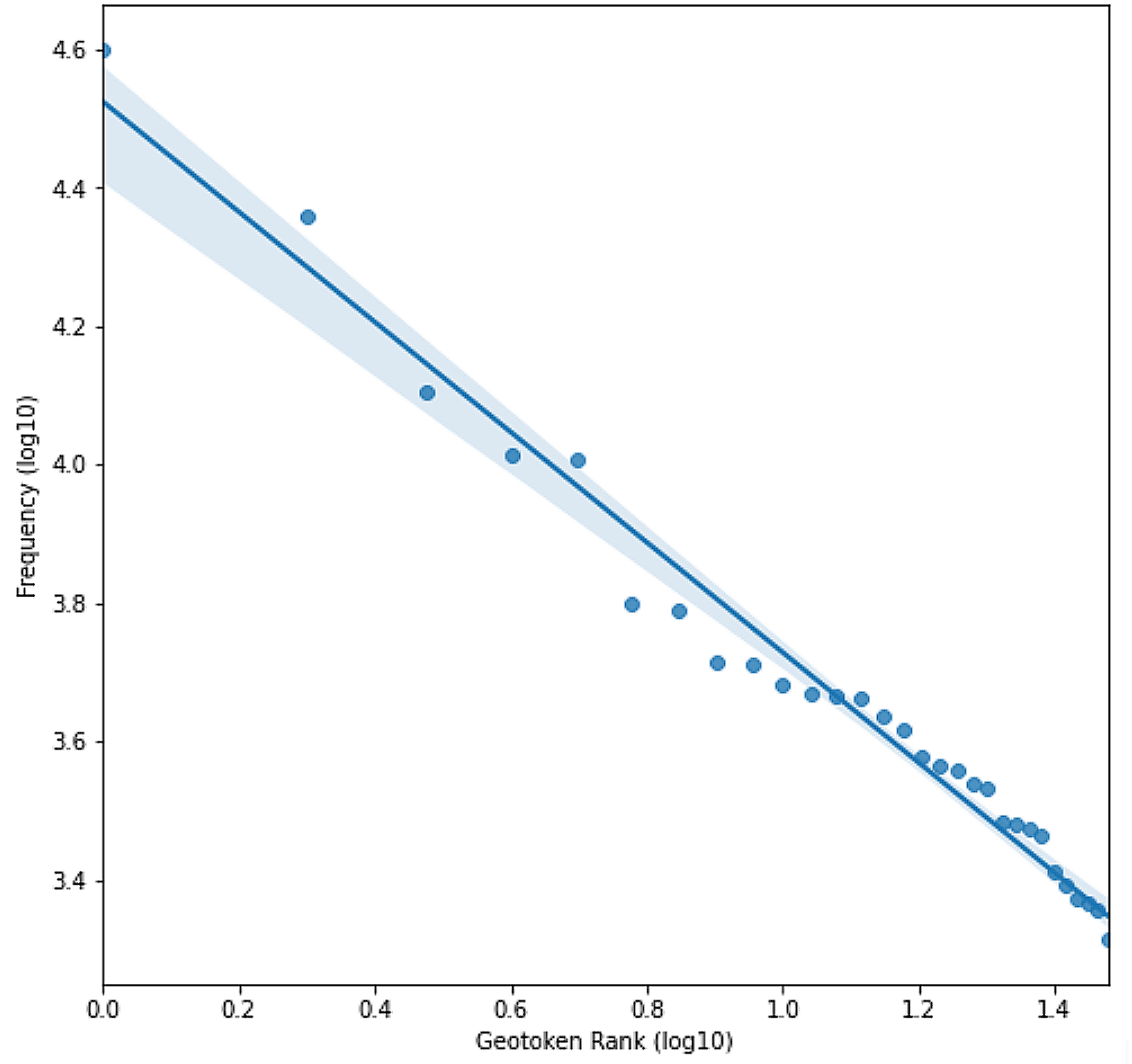}
  \caption{A linear relationship between the log rank of geo-tokens and their log frequency.}
  \label{loggeo}
\end{figure}

\begin{table}[hbt!]
 \caption{Random examples of geographical tokens intransaction messages}
  \centering
  \begin{tabular}{lll}
    \toprule{}
    Token     & Examples \\
    \midrule
    Las Vegas & Las Vegas Hotel \\
            & Accommodations in Las Vegas    \\
            & Vietnamese food from Las Vegas \\
            & Las Vegas \\
    \midrule
    Tahoe & Tahoe car + gas \\
            & Tahoe    \\
            & Tahoe groceries \\
            & Tahoe groceries nom nom \\
    \midrule
    Coachella & Coachella snacks \\
            & Coachella ’16    \\
            & Coachella \\
            & Coachella lockers \\
    \midrule
    China & China virus \\
            & China town    \\
            & China food \\
            & China Trip \\
    \bottomrule
  \end{tabular}
  \label{geographical_tokens}
\end{table}

\hfill{}\break{}
\hfill{}\break{}
\hfill{}\break{}
\hfill{}\break{}
\hfill{}\break{}
\hfill{}\break{}
\hfill{}\break{}
\hfill{}\break{}
\hfill{}\break{}
\hfill{}\break{}

\begin{figure}[hbt!]
  \centering
  \includegraphics[width=0.9\columnwidth]{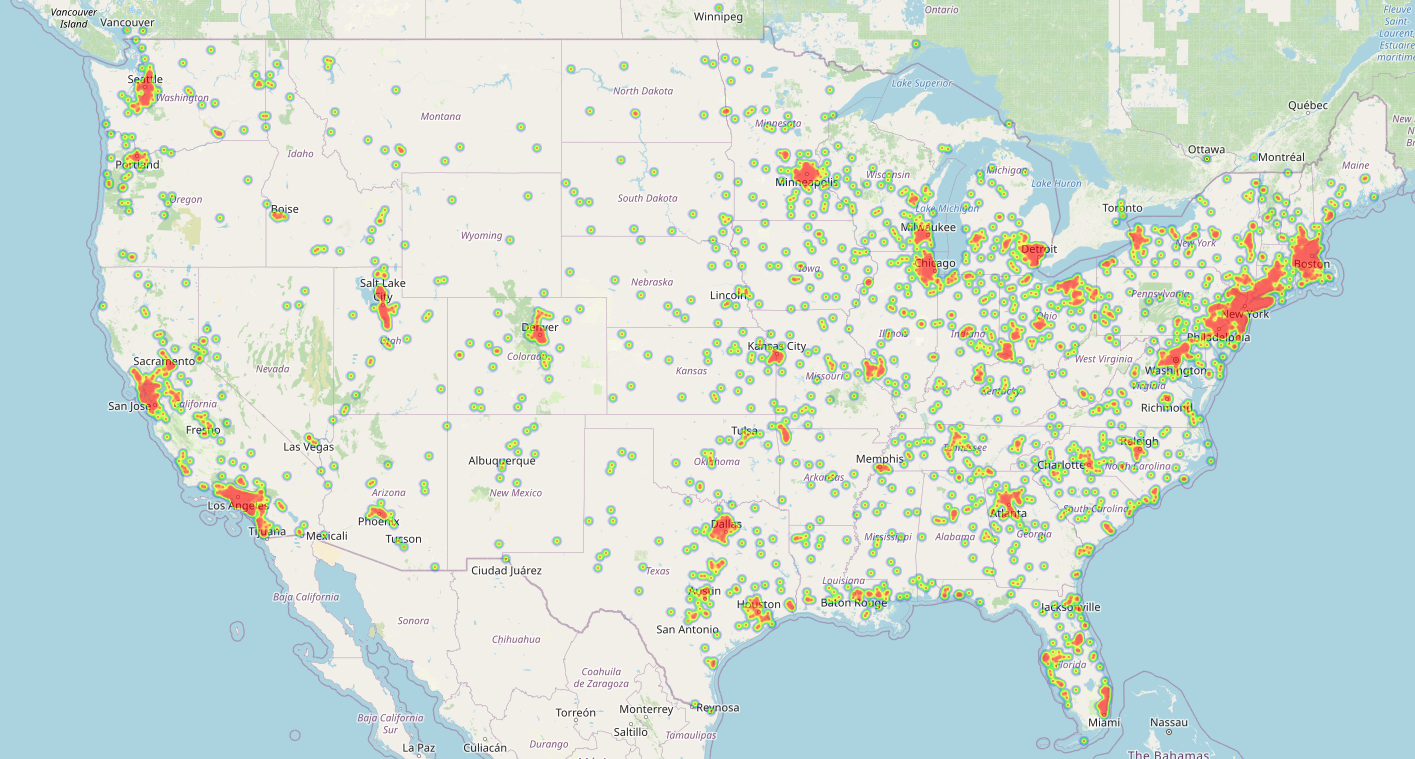}
  \caption{Locations of Venmo users according to geotags on paired Facebook profiles.}
  \label{heatmap}
\end{figure}

Figure \ref{heatmap} illustrates the locations of the 7K users who we were able to geolocate using Facebook geotags. This represents 6K more than used in Yao \cite{yao2018beware}. As expected, we see more users in populous areas with a correlation of 0.93 between the population of a state and the number of users found geotagged there. (We used the Pearson method to estimate the correlations between the use of COVID-19 keywords in a region and the number of reported COVID-19 cases.) This is in stark contrast to the (less accurate) results of Figure \ref{mordecai}, which shows the majority of users being in Las Vegas and Tahoe. We were able to find significantly more profiles using Bing. However, since the results did not contain profile pictures, it was impossible to disambiguate multiple Facebook profile results for a given name. This occurred even when the name was unique among the 22.1M users in our dataset and because of this our geotag results are solely based on profiles discovered on PeekYou and the text found on the accompanying Facebook profiles.

\subsection{COVID-19 Keywords}
To extend the work of others done using Twitter data \cite{bisanzio2020use}, we measured the frequency of COVID-19 related keywords in the memo field of Venmo transactions. We use the keywords “covid”, “coronavirus”, “isolating”, “fever”, “cough”, “symptoms”, “pneumonia” (borrowed from \cite{mackey2020machine}) as well as our own terms of “quarantine”, “social distancing”, and “self isolating”. We find an average of 1 COVID-19 related keyword per 636 transactions in the dataset. As Table \ref{transaction_message} shows, these terms are used in several different contexts including discussing one’s own behavior (“quarantine”) and symptoms/treatment (“cough syrup”), discussing others (“PEOPLE are bad at social distancing”), describing pandemic altered activities (“fantasy covid football”\footnote{Fantasy football is a common use of Venmo \cite{acker2020venmo}\cite{zhang2017cold}}), as well as self-described risky behaviors (“Risking covid for a wine and dine”). On manual inspection of our collected data, we found that the use of symptom-related keywords such as “fever”, “pneumonia”, and “cough” appear to be related to the condition of the posting user. Since the messages are typically related to the transaction taking place, it will be less common to have a transaction that says “I have covid” than for a transaction to say “fever medicine” for a user who actually has COVID-19, although both do occur in the dataset.

\begin{table}[hbt!]
 \caption{Examples of COVID-19 keywords in transaction messages}
  \centering
  \begin{tabular}{lll}
    \toprule{}
    Token     & Examples \\
    \midrule
    social distancing & i $\heartsuit$ social distancing \\
            & social distancing    \\
            & PEOPLE are bad at social distancing \\
            & practice social distancing \\
    \midrule
    quarantine & quarantine shenanigans \\
            & quarantine shenanigans    \\
            & week 87 of quarantine \\
            & We will be under quarantine \\
    \midrule
    covid & covid-19 \\
            & Risking covid for a wine and dine    \\
            & it’s the covid test for me \\
            & fantasy covid football \\
    \midrule
    cough & Cough cough \\
            & cough syrup and sprite    \\
            & Eggs and cough syrup \\
            & cough he \\
    \bottomrule
  \end{tabular}
  \label{transaction_message}
\end{table}

\begin{figure}[hbt!]
  \centering
  \includegraphics[width=1\columnwidth]{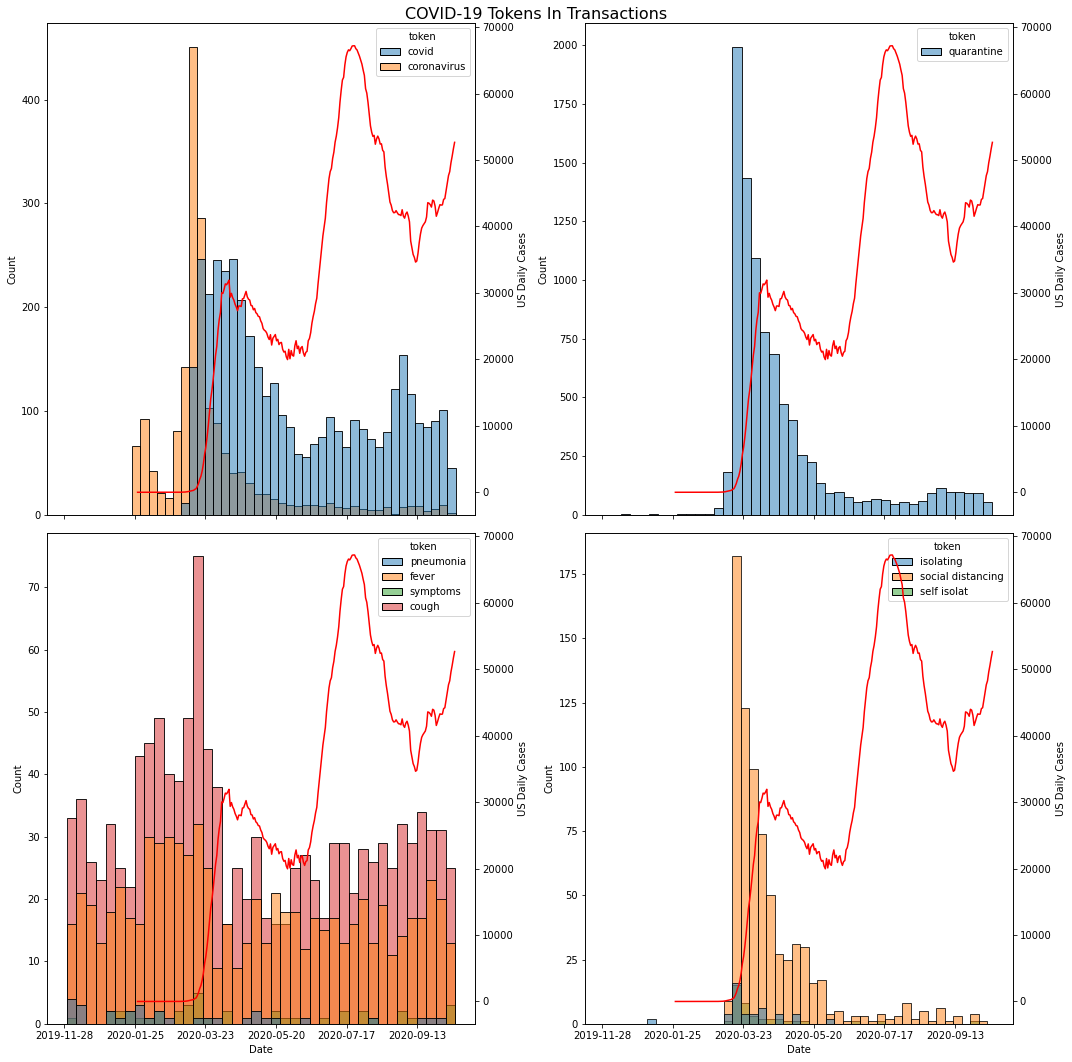}
  \caption{The frequency of several COVID-19 keywords over time; U.S. reported COVID-19 cases are overlaid in red.}
  \label{token_transactions}
\end{figure}

Figure \ref{token_transactions} illustrates the use of COVID-19 keywords over time and indicates a significant spike in the usage of “quarantine”, “covid”, “coronavirus”, “social distancing”, “isolating”, “cough”, and, although not as significant, “fever” in March 2020. This time frame is aligned with the U.S.’s first surge of reported COVID-19 cases. Of these, the spike in use of “quarantine” is the most significant with a peak usage higher than that of all other terms combined. A jump is observed in both “cough” and “fever” as early as late January which could be a culmination of flu season and possibly unidentified COVID-19 cases. Surprisingly, the second surge of cases in July 2020 does not result in a similar spike and the use of “symptoms” and “pneumonia” remain relatively constant and infrequent throughout the timespan. In addition, there is a small increase in the use of “covid”, “cough”, and “quarantine” in September 2020 between the 2nd and 3rd case peaks where cases actually drop. We also find that while “coronavirus” is initially more popular than “covid”, from April 2020 onwards, “covid” becomes more commonly used over time, mirroring general usage.

\begin{table}[hbt!]
 \caption{Correlation values; *** $= P<0.001$, ** $= 0.001<P<0.01$, * $= 0.01<P<0.1$}
\begin{tabular}{p{30pt}p{30pt}p{26pt}p{24pt}p{24pt}p{24pt}p{20pt}p{27pt}p{27pt}p{30pt}p{25pt}p{30pt}p{24pt}p{26pt}}
\toprule{}
                     & pop-ulation & covid cases & cluster counts & covid counts & sum size & sum trans & avg cluster size & avg cluster trans & covidtok /cluster & covidtok /pop & cases /pop & clusters / pop & covidtok / cases \\
\midrule{}
pop-ulation           & 1          &             &              &            &          &           &                &                 &                      &                  &               &                  &                    \\
\midrule{}
covid cases          & 0.92***    & 1           &              &            &          &           &                &                 &                      &                  &               &                  &                    \\
\midrule{}
cluster counts         & 0.78***    & 0.82***     & 1            &            &          &           &                &                 &                      &                  &               &                  &                    \\
\midrule{}
covid counts           & 0.77***    & 0.81***     & 1***         & 1          &          &           &                &                 &                      &                  &               &                  &                    \\
\midrule{}
sum size             & 0.78***    & 0.81***     & 1***         & 1***       & 1        &           &                &                 &                      &                  &               &                  &                    \\
\midrule{}
sum trans            & 0.78***    & 0.81***     & 1***         & 1***       & 1***     & 1         &                &                 &                      &                  &               &                  &                    \\
\midrule{}
avg cluster size       & -0.03      & -0.02       & -0.07        & -0.05      & -0.03    & -0.02     & 1              &                 &                      &                  &               &                  &                    \\
\midrule{}
avg cluster trans      & -0.01      & -0.02       & -0.05        & -0.04      & -0.02    & 0         & 0.93***        & 1               &                      &                  &               &                  &                    \\
\midrule{}
covidtok /cluster & -0.09      & -0.04       & -0.01        & 0.06       & 0        & 0         & 0.12           & 0.06            & 1                    &                  &               &                  &                    \\
\midrule{}
covidtok /pop     & -0.24      & -0.2        & 0.09         & 0.13       & 0.08     & 0.08      & -0.16          & -0.11           & 0.27*                & 1                &               &                  &                    \\
\midrule{}
cases /pop        & 0.02       & 0.28*       & 0.12         & 0.14       & 0.13     & 0.13      & 0.11           & 0.07            & 0.25*                & -0.05            & 1             &                  &                    \\
\midrule{}
clusters /pop     & -0.24*     & -0.2        & 0.11         & 0.13       & 0.09     & 0.09      & -0.19          & -0.12           & 0.12                 & 0.98***          & -0.09         & 1                &                    \\
\midrule{}
covidtok /cases   & -0.27*     & -0.27*      & -0.02        & 0          & -0.03    & -0.03     & -0.15          & -0.11           & 0.05                 & 0.74***          & -0.42**       & 0.75***          & 1           
            \\
\bottomrule{}
\end{tabular}
\end{table}

\begin{figure}[hbt!]
  \centering
  \includegraphics[width=0.6\columnwidth]{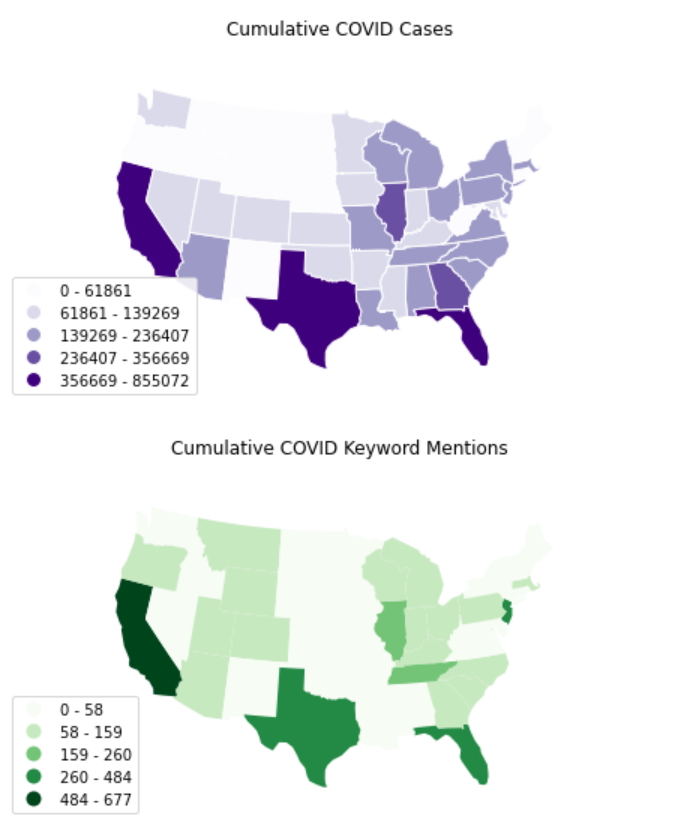}
  \caption{a) Cumulative number of COVID-19 cases by mainland US State and b) cumulative number of COVID-19 keywords mentioned in our data by mainland US State.}
  \label{cum_state}
\end{figure}

\hfill{}\break{}
\hfill{}\break{}
\hfill{}\break{}
\hfill{}\break{}
\hfill{}\break{}
\hfill{}\break{}
\hfill{}\break{}
\hfill{}\break{}
\hfill{}\break{}
\hfill{}\break{}
\hfill{}\break{}
\hfill{}\break{}
\hfill{}\break{}
\hfill{}\break{}
\hfill{}\break{}
\hfill{}\break{}
\hfill{}\break{}
\hfill{}\break{}
\hfill{}\break{}
\hfill{}\break{}
\hfill{}\break{}
\hfill{}\break{}
\hfill{}\break{}
\hfill{}\break{}
\hfill{}\break{}
\hfill{}\break{}
\hfill{}\break{}
\hfill{}\break{}
\hfill{}\break{}
\hfill{}\break{}
\hfill{}\break{}
\hfill{}\break{}
\hfill{}\break{}\hfill{}\break{}
\hfill{}\break{}
\hfill{}\break{}
\hfill{}\break{}
\hfill{}\break{}
\hfill{}\break{}
\hfill{}\break{}

\subsection{Location-Based Clusters}
Using all the transactions in our dataset (March - October 2020), we construct a graph of users connected by their transactions. To prevent extremely large groups created by connecting users who don’t often use Venmo to transact with each other, we remove edges with less than 4 transactions. We then use connected components to obtain 902K distinct groups with a mean size of 10.2 users. Of these, we use the locations of geotagged users or, if there are no geotagged users, the location of the most common geolocation-related token referenced in their transactions. Discarding groups with no clear location, we obtain 38K geo-located Venmo groups. Applying a user’s geolocated group as an estimate for their location, we found a correlation of 0.78 between the population of a state and the number of users estimated to be located there. This signifies a decrease in accuracy over using only geotagged users (which had a correlation of 0.93), but allows us to increase our number of geolocated users by a factor of 5. Figure \ref{cum_state} illustrates the cumulative use COVID-19 keywords by these groups and the relationship between these keywords and cumulative cases in U.S. states. While there is a strong linear relationship (correlation of 0.81) between COVID-19 keywords and cases, it should be noted that this is still less than the correlation between cumulative cases and the population of each state (0.91). We also find that of groups that mention a COVID-19 keyword, in 78\% of occurrences that was the only time a keyword was mentioned in the group.\\
\begin{figure}[hbt!]
  \centering
  \includegraphics[width=0.9\columnwidth]{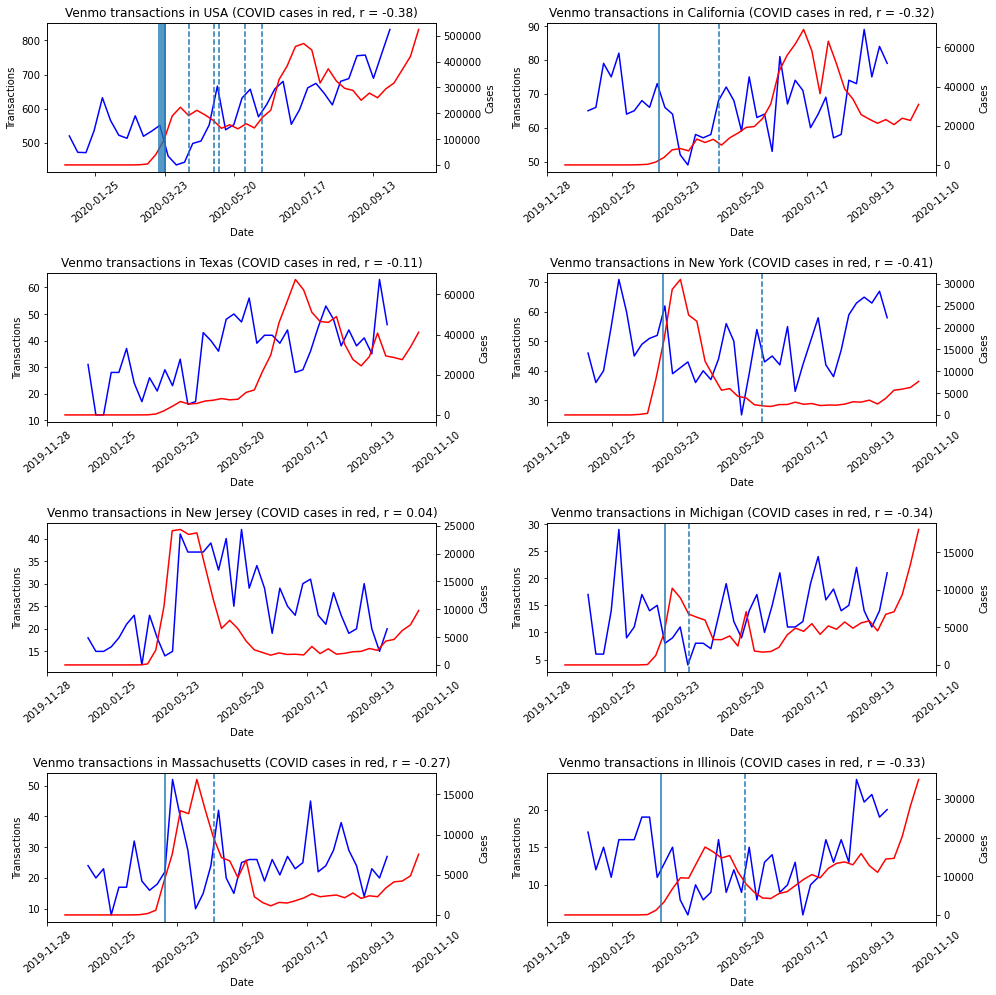}
  \caption{The frequency of Venmo transactions over time in a given region; regional reported COVID-19 cases are overlaid in red and state lockdowns are shown as vertical lines (a dashed line represents a lockdown ending).}
  \label{state_transactions}
\end{figure}
\hfill{}\break{}
\hfill{}\break{}
\hfill{}\break{}
\hfill{}\break{}
\hfill{}\break{}
\hfill{}\break{}
\hfill{}\break{}
\hfill{}\break{}
Figure \ref{state_transactions} illustrates the relationship between COVID-19 cases and the number of transactions involving verified groups (groups that have at least one geotagged member) in a given region is illustrated. We found a weak negative correlation between reported cases and number of transactions in most states with a correlation of -0.49 in the U.S. overall. For the U.S. as a whole, the lowest rate of Venmo transactions (and, inversely, an increase in COVID-19 cases) are seen after several state pandemic lockdowns are put into effect. We then see a steady increase in transactions as those lockdowns are lifted. Looking at state-specific charts (chosen for having a large number of datapoints), the effects of lockdowns and relationship between cases and transactions is less clear. While it seems in some cases the number of transactions drop with the introduction of lockdowns, the high variance of transaction frequencies over time (at least partially due to our limited number of geotagged users in each location), prevents this conclusion from being significant. Peak case counts tended to occur within the lockdown window for several states, however, this is not observed in Venmo transactions nor in the US as a whole. New Jersey and Illinois both show a trend of decreasing transactions during the later half of 2020 but increasing cases.

\section{Discussion}
\subsection{Data Collection}
In this study, we successfully collected Venmo data at scale and used big data, machine-learned, and statistical methods to cluster users who may have had physical contact. For each of these clusters, we examined their use of COVID-19 related keywords and attempted to localize them using both location-related keywords and Facebook geotags. It should be noted that Venmo, as a platform, has traditionally been biased towards primarily English-language speaking young adults in the U.S \cite{caraway2017friends}. This should be taken into consideration when using Venmo (or any other social media platform) for infoveillance as other demographic groups will be severely underrepresented. Since neither the Venmo API nor Facebook profiles yielded this kind of demographic information, we were unable to further study these implications and we do believe this is an important area for future research.

It should also be noted that significant effort was required to obtain our dataset, highlighting larger issues with social media platforms restricting or blocking API access to publicly available data. While studies such as Mackey et al. \cite{mackey2020machine} use geo-located data from a formal API provided by Twitter, using Venmo for similar lines of research presents several new challenges. First, Venmo’s REST API is undocumented and its endpoints and required parameters are only known through reverse engineering the Venmo app. Of these endpoints, only “GET /api/v5/users/:userId/feed” (according to our experience during this study) provides historical transaction data which therefore requires us to guess and query user IDs in order to populate our dataset. This is in comparison to Twitter’s documented and extensive query options\footnote{\href{https://developer.twitter.com/en/docs/api-reference-index}{https://developer.twitter.com/en/docs/api-reference-index}}. Second, although Venmo has a controversial public-by-default policy \cite{paul_2018, tandon2022know}, some do not feel comfortable having payment transactions on Venmo be public, opting to enable privacy conscious settings \cite{caraway2017friends}, which hides their results from API responses. Third, geolocation studies of Venmo have the added difficulty of Venmo not explicitly collecting location data. This forces our study as well as others to rely on Facebook-integrated user profiles as a proxy for collecting geotag information which reduces the size of one’s dataset by approximately 90\%. Though Facebook does have an API, it does not have the ability to query by publicly available profile picture nor does it provide a method of querying a user’s geotags. This forces the data collection to be done through browser automation, which is prone to breaking upon site changes. Lastly, the rate limiting imposed by Venmo during the end of our data collection, if permanent, would make collecting a dataset of our size (136M transactions) impossible in future studies. The rate limiting seems to be non-specific to our account and we hope Venmo provides an alternative mechanism for collecting data to aid future social research. While these issues with Venmo hinder research of the platform, data availability changes may stem from Venmo seeking to protect the data of its users. If this is the case, a formal and regulated API by Venmo - like that of other major social media platforms like Facebook and Twitter - would be a solution. These are issues not just specific to Venmo, but will be faced by researchers collecting data from other, niche social media platforms. For these reasons it is critical that social media researchers responsibly share their datasets to aid future progress in this field. Shared datasets could use hashing or other measures to protect personally identifiable information (PII).

Using Facebook as a proxy for geolocation also created independent challenges, mainly due to rate limiting and other measures. After only a few hundred requests, Facebook would temporarily lock our account. While Facebook prevents us from querying users directly, PeekYou’s mirror of the same information and lack of any authentication or rate limiting, allowed us to perform Venmo-Facebook profile matching. We were also able to query PeekYou using their API without using browser automation. While this enables us as researchers to study a larger sample, we would have preferred to directly query data from Facebook as users may change their profile information, make their data private, etc. Mirrors of publicly available social media data such as PeekYou could have out of date information or even display profile information for a user who had since changed their privacy settings on Facebook. This highlights the importance of social media platforms providing academic researchers with higher levels of API access to public data given the utility of these data to questions in domains such as public health.

\subsection{User Location}
Though we were able to successfully infer user location via social payments, we did encounter challenges as our geoparsing of Venmo memo messages often identified locations when a user was away from their home location. This does not necessarily make localizing groups using keywords inaccurate. If money is sent between users for “Las Vegas rental car” it can be estimated a user is likely in Las Vegas. However, this is merely an estimation. Moreover, we found that users tend to mention location in their transaction when the location is travel-related and not their home location. A transaction sent by a user who went to a restaurant in their hometown/local area will likely just mention “restaurant” whereas if the transaction took place on a trip, we found that they may also share the location as part of the description (e.g, “las vegas restaurant”). To help ameliorate this effect, we also collected geotags for several thousand users in our dataset. By looking at the correlation between our predicted users’ geolocations and the actual population of those locations, we confirm this to be more accurate and rely on it to enhance our geolocation estimates of clusters who have geotag data. The use of geotags over geo-parsing is also evident in Twitter research from the datasets used by others. Future work would benefit from further methods to derive geotags from social platforms that do not provide them through readily accessible APIs.

\subsection{COVID-19 Keywords}
We found that COVID-19 related keywords do appear within mobile social payments on Venmo. We found a significant spike in the use of the word “quarantine” and keywords in March 2020 which follows Twitter analysis of symptoms during the same timespan by others (e.g., \cite{guo2020mining}, \cite{chandrasekaran2020topics}). We did not see a similar spike in the latter part of 2020 (e.g, during the second surge of reported U.S. COVID-19 cases). One explanation could be that as a result of social distancing and self-isolation, users stopped or reduced meeting in-person and thus Venmo use declined. However, this is not supported by the results of Figure \ref{state_transactions} which shows an increase in transactions during that time frame. This is further supported by marked increases in COVID-19 cases in the US \cite{shang2022global}. Another explanation is that the March 2020 spike corresponded to several U.S. state and local lockdowns going into effect. This was a specific time when there was significant user and media attention to the pandemic \cite{mach2021news}, which reduced as users acclimatized to the situation. Looking at the use of symptom keywords in Figure \ref{token_transactions}, we see an increase in the use of “fever” and “cough” preceding the initial spike of reported cases. We believe this is an illustration of flu season combined with unreported COVID-19 cases.

\subsection{Location-Based Clusters}
Like others, we found that constructing a social graph is a useful way to leverage Venmo as a social platform that is often utilized for in-person transactions. Creating an interaction graph also allowed for further analysis in simulations and epidemiological modeling such as in Farrahi et al \cite{farrahi2014epidemic}. We created our graph by making each user a vertex and connecting users if they have had a transaction between them. We then filter the edges to only those that have a multiplicity of at least 4 from January to October 2020. This is necessary to avoid creating million-user groups and to strengthen the assumption that there is some sort of relationship between the users during the time period of our study. The results can also be somewhat confounded by the presence of same-location relationships (e.g., Venmo transactions between friends who regularly meetup and live in the same area) vs remote relationships (e.g., Venmo transactions between users who live in separate regions but are visiting each other during their transactions) since we filter by the number of transactions rather than where they took place. When we plot transactions done by these groups over time in Figure \ref{state_transactions}, results are as expected in aggregate, but not regionally. This could be due to the inaccuracies in our method of geolocation and/or the regionally varying use of Venmo as users adapt to the pandemic and move to a higher frequency of remote transactions.

\section{Future Work and Limitations}
The study of social media data in pandemic-related infoveillance has been growing and we expect more people will share relevant information on various online social platforms in the future. Most social media-based infoveilance studies rely on data from Twitter, which may poorly correlate with in-person contact or location-based interactions. Future studies would benefit from exploring social platforms such as Venmo, where online interactions correlate with in-person ones and take advantage of richer social media post formats. For example, tagged pictures could indicate that all the users tagged were physically together when the picture was taken or social media-based event ticketing for in-person events could identify large groups where viral transmission could have occurred. With social media companies taking measures to limit API and third-party access, these studies would likely also benefit from a partnership with the social media companies themselves. Significant data collection and analysis is already done by these companies and it would be beneficial if a similar effort could be applied to empower contact tracing, which has the potential to save lives in a pandemic like COVID-19 \cite{firth2020using}. While platforms such as Facebook can already estimate in-person contact with automated photo tagging\footnote{\href{https://www.cnet.com/news/facebook-faces-fresh-privacy-palaver-over-face-recognition-for-photo-tagging/}{https://www.cnet.com/news/facebook-faces-fresh-privacy-palaver-over-face-recognition-for-photo-tagging/}}, social features such as setting an “infection status” and alerting individuals that they may have been near someone who tested positive do not yet exist. They would, however, be relatively easy to implement as they extend from similar features on Facebook such as a user marking themself as safe during a disaster using the platform’s ‘Safety Check’ feature \cite{pandey2017safety}.

Our study marks the first time Venmo-derived data was used to construct a viable social graph for epidemiological infoveillance. Future work should look into more accurate methods for deriving contact-relevant user metadata such as geolocation on platforms that do not provide these data (such as Venmo). It may be possible to use a machine learning-based analysis to derive further insights from transaction messages. Future work could also compare these methods with large scale ground truth contact tracing data. Moreover, additional research could explore how a graph derived from social media data can be extended into simulation models (i.e., building on the methods developed by Farrahi et al. \cite{farrahi2014epidemic} \cite{rusu2021modelling}).

It is clear that the viability of social media-based infoveillance will heavily depend on both location accuracy and on the population that uses the platform the data is being derived from. Moreover, “early fintech adopters tend to be tech-savvy, younger, urban, and higher-income individuals \cite{lee2018fintech}.” Because of this, the process of tracing people outside of particular demographic groups could be limited if Venmo is heavily skewed. For this reason, future work should study ways of augmenting existing methods of infoveillance which are more representative of the general population with specialized insights derived from social media analysis. While infoveillance naturally benefits from more data, social media-based trace data would also help enhance the datasets used by epidemiologists and policymakers. Future work could also look further into specific applications in places such as cities, interconnected communities, and college campuses for example.
Our study was most limited by data collection restrictions. While we were initially able to collect several million Venmo transactions quickly, changes in Venmo’s API forced us to truncate our dataset to only a quarter of total users. Moreover, Facebook’s rate limiting meant that we could only tie a small fraction of these users to physical locations for analysis. These were limitations beyond our control. 

\section{Conclusion}
Social media-based infoveillance is an important modality to limiting disease spread and ultimately saving lives. While data collection infrastructure has been deployed in a variety of forms globally, distrust of both public and private solutions (e.g., contact tracing apps) and substantial privacy concerns have prevented them from becoming widely adopted in countries like the U.S. (where a disproportionate number of COVID-19 deaths occurred \cite{shang2022global}). It is critical that a privacy-utility balance is found to benefit future data collection methods. Our study finds that using publicly available social media data on novel platforms could be an infoveillance solution in countries such as the U.S., where users are open to sharing their location and transactional information publicly, but may not be open to using dedicated contact tracing apps and platforms. Unlike most existing studies in this area which use Twitter data, we uniquely used sparse, but rich data from the U.S.-based social payments platform, Venmo, which we were able to successfully geoparse. 
In this study, we collected transactions and user data from Venmo’s public API endpoint. We then used PeekYou to query users by name and identify matching Facebook profiles using profile picture similarity. With these profiles, we used public Facebook profiles to extract geotag information. We were able to collect data for 22.1M users, of which 19K are tied to a Facebook profile and 7K are geolocated. Since this is only a small portion of the dataset, geoparsing is also employed on transaction messages. With our dataset of 136M transactions, we created location-based clusters connecting users who have had multiple transactions between them. Connected components, geotagged users, and transaction geoparsing are then combined to obtain 38K geolocated clusters which show differences in COVID-19 keyword use at different locations which vary over time. We found a strong correlation between the use of COVID-19 keywords in these groups and COVID-19 cases in mainland U.S. states. In aggregate, lockdowns correlated with the frequency of transactions with COVID-19-related content in these groups. We also studied the usage of several COVID-19 keywords over time in the memo field of Venmo transactions and found significant spikes in their usage during March 2020. This is also seen in the usage of symptom-specific keywords such as “fever” and “cough”.
Our data sourcing method of using Venmo is the first of its kind. We hope that future work is able to use public social media data from smaller social media platforms alongside data from the traditional sources of Twitter and Facebook. We believe that these types of triangulated methods will yield more robust location information, an attribute particularly critical to any tracing application. Moreover, unlike Facebook and Twitter, Venmo is often used by users as an extension of in-person meetups. Our study ultimately finds that using public data collected from Venmo provides value for surviving viral outbreaks. Despite the challenges of using Venmo as a data source, we were able to collect a large dataset of Venmo transactions, verify user location information, and analyze these data in the context of the COVID-19 pandemic. Our intention is that these methods help pave the way for future research into extending existing pandemic tracking methods, which leverage more niche public social media data.

\section{Source Code}
The source code used for generating the dataset, undertaking our research, and rendering visualizations is available at \href{https://github.com/sshh12/venmo-research}{https://github.com/sshh12/venmo-research}.

\section*{Acknowledgments}
We thank Pranav Venkatesh [\texttt{pranav.venkatesh@utexas.edu}] for his help formatting plots, editing, and migrating our manuscript to arXiv.

%Bibliography
\bibliographystyle{unsrt}  
\bibliography{references}

\end{document}